# Tracking valley topology with synthetic Weyl paths

Xiying Fan, Tianzhi Xia, Huahui Qiu, Qicheng Zhang, and Chunyin Qiu[*]

Key Laboratory of Artificial Micro- and Nano-Structures of Ministry of Education and School of Physics and Technology

Wuhan University, Wuhan 430072, China

[*] To whom correspondence should be addressed: cyqiu@whu.edu.cn

*Abstract*. Inspired by the newly emergent valleytronics, great interest has been attracted to the topological valley transport in classical metacrystals. The presence of nontrivial domain-wall states is interpreted with a concept of valley Chern number, which is well defined only in the limit of small bandgap. Here, we propose a new visual angle to track the intricate valley topology in classical systems. Benefiting from the controllability of our acoustic metacrystals, we construct Weyl points in synthetic three-dimensional momentum space through introducing an extra structural parameter (rotation angle here). As such, the two-dimensional valley-projected band topology can be tracked with the strictly quantized topological charge in three-dimensional Weyl crystal, which features open surface arcs connecting the synthetic Weyl points and gapless chiral surface states along specific Weyl paths. All theoretical predictions are conclusively identified by our acoustic experiments. Our findings may promote the development of topological valley physics, which is less well-defined yet under hot debate in multiple physical disciplines.

*Introduction.* Valley electrons in two-dimensional (2D) materials have sparked extensive attention in multidisciplinary fields [1-16]. The discrete valley index, labeling the degenerate energy extrema in the band structure, can be viewed as a new controllable carrier of quantum information when the intervalley scattering is negligibly weak, much like the spin in spintronics. Many fascinating phenomena associated with valley-contrasting electronic properties have been reported [1-16], such as valley filtering and topological valley Hall effect, which are paving the road for designing novel modern electronic devices.

Inspired by the concept of valleytronics, the peculiar valley physics has been migrated to metacrystals for classical waves [17-44]. Besides the valley-locked bulk transport [17,18,23,24], the marriage of the exotic valley degree of freedom and metacrystals provides a simple yet highly efficient recipe for realizing topological transport of classical waves at the interface of distinct valley Hall phases [19-44]. Markedly different from the traditional waveguide states [45-47], the valley-projected interface states can serve as the basis of designing integrated devices with unconventional functions, given many exceptional transport properties such as negligible backscattering to sharp bending corners [19-44].

Valley Chern number (VCN), $C_v$, defined with an integral of Berry curvature in individual valleys, serves as a key physical quantity that universally characterizes the topological valley transport in quantum [6-16] and classical [19-44] systems, despite of rather different mechanisms introduced for valley Hall phase transition. For the widely explored domain-wall system formed by topologically distinct valley Hall phases, the magnitude of their VCN difference across the domain-wall, $|\Delta C_v|$, gives the number of valley-locked interface modes. In most classical systems, $C_v = \pm 1/2$ and $|\Delta C_v| = 1$ [19-44], accounting for one pair of one-dimensional (1D) time-reversal-related interface modes. Nonetheless, an unambiguous VCN definition relies on extremely localized Berry curvature in each valley [2], and $|\Delta C_v|$ approaches 1 only in the limit of narrow bandgap; the deviation from 1 becomes serious for a system with wide bandgap (which is highly-desired in real applications [20]). Therefore, the band topology of such domain-wall systems is subtle since there is no strict topological invariant to establish bulk-boundary correspondence. Even worse, for a single valley metacrystal positioned in a wave-impenetrable uniform media ($C_v = 0$), there is no any deterministic knowledge on the presence of edge states, since the VCN difference $|\Delta C_v| \simeq 1/2$ does not contribute a (even approximate) bulk-boundary correspondence.

Here, we unveil the elusive valley-projected band topology from a perspective inherent in three-dimensional (3D) Weyl physics [48-59]. We consider acoustic valley metacrystals (AVMs) and extensions can be made to the other classical systems straightly. Our AVM consists of rotable triangular scatterers in a triangular lattice, in which valley Hall phase transition can be realized by simply rotating the anisotropic scatterers [Fig. 1(a)]. Weyl points (WPs), isolated linearly crossing points in 3D band structures, are synthesized in a virtual 3D momentum space spanned by the physical dimensions plus the scatterer's orientation degree of freedom [Fig. 1(b)]. Resorting to the bulk-boundary correspondence established by the strictly quantized WP charge, gapless 'surface' states, which emerge at the 1D edges of individual AVMs, are predicted along specific Weyl paths in the synthetic surface Brillouin zone (BZ), as a faithful manifestation of the global band topology for the parameter-tunable AVMs. The physical interpretation is extended to the more popular domain-wall system [19-44], which is synthesized as a stacking of two oppositely-charged Weyl crystals. All theoretical predictions are unambiguously validated by our acoustic experiments through measuring 1D edge/interface states.

*AVMs and synthetic WPs.* As illustrated in Fig. 1(a), our AVM is constructed by a triangular lattice of rigid triangular scatterers in air background. The lattice constant $a = 17.2$ mm, and the volume filling ratio of the scatterer $\eta \simeq 0.243$. The orientation of the scatterer, characterized by the angle $\theta$ with respect to $y$ axis, enables a flexible control over the spatial symmetry and bandgap of the system. We only consider $\theta \in [-60°, 60°]$, a rotation period of the scatterer. Specifically, for $\theta = \pm 30°$, the point group at the BZ corner K(K') features $C_{3v}$ symmetry owing to the perfect match of the scatterer's mirrors to those of the triangular lattice. Apart from those special angles, the mirrors are mismatched and the crystal symmetry reduces to $C_3$. As such, the 2D band structure of the AVM supports a twofold linear degeneracy at K(K') for $\theta = \pm 30°$, whereas it is lifted for a generic $\theta$. This is exemplified by the cases of $\theta = 30°$ and $15°$ [Fig. (1c)]. For brevity, throughout this work the dimensionless frequency $a/\lambda$ is employed in the band structure, with $\lambda$ being sound wavelength in air.

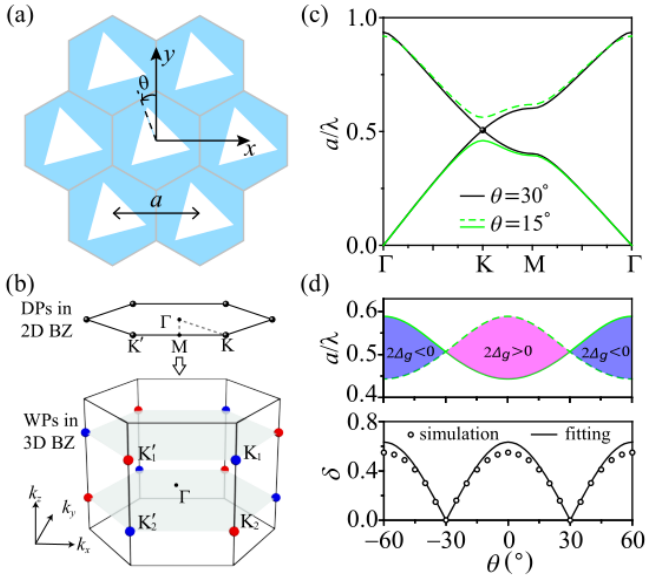

FIG. 1. AVMs and synthetic WPs. (a) 2D AVM made of regular triangular scatterers (white) in air background (blue). The rotation angle $\theta$ defines the orientation of the anisotropic scatterers. (b) 2D BZ and its 3D extension (with $k_z$ mimicked by $\theta$). The DPs (black spheres) in 2D evolve into WPs (color spheres) in the synthetic 3D BZ, where red and blue label their topological charges of $\pm 1$. (c) 2D band structures for the AVMs of $\theta = 15°$ and $30°$. (d) Top panel: $\theta$-evolution of the band-edge frequencies at K, where the sign of bandgap ($2\Delta_g$) distinguishes topologically distinct AVH phases. Bottom panel: relative deviation of the Berry phase ($\delta$) plotted as a function of $\theta$, which becomes notable in the case of big bandgap.

Below we focus on the K valley and the physics of K' valley can be inferred by time reversal. Notice that the bandgap closes linearly as $\theta$ approaches $\pm 30°$, as shown in the top panel of Fig. 1(d) by the angularly-dependent band edge frequencies at K. This suggests a signal of acoustic valley Hall (AVH) phase transition [20], which can be captured by a $\theta$-dependent effective Hamiltonian derived from $k \cdot p$ perturbation method,
$$\delta H_K(\theta) = v_D q_x \sigma_x + v_D q_y \sigma_y + \Delta_g(\theta) \sigma_z. \quad (1)$$
Here $v_D$ is Dirac velocity of the AVM with $\theta = \pm 30°$, $\mathbf{q} = (q_x, q_y)$ measures the momentum deviation from K, and $\sigma_i$ are Pauli matrices. The sign of the bandgap $2\Delta_g$, determined by the frequency order of the oppositely rotated valley vortex states [20], characterizes two distinct AVH insulators. The effective Hamiltonian in Eq. (1) gives a localized Berry curvature around K for the lowest band [2],
$$\Omega(\mathbf{q}) = \frac{1}{2} \Delta_g v_D^2 (v_D^2 q^2 + \Delta_g^2)^{-3/2}. \quad (2)$$
Its integral over the K valley yields a Berry phase $\pi \text{sgn}(\Delta_g)[1-\delta]$, where the dimensionless quantity $\delta = v_D^{-1} q_c^{-1} |\Delta_g| > 0$ characterizes the relative deviation of Berry phase from the quantized value, $\pi \text{sgn}(\Delta_g)$, and $q_c$ is a momentum truncation resulted by the finite-sized BZ. In the limit of zero bandgap, the VCN for the K valley $C_K = \frac{1}{2} \text{sgn}(\Delta_g)$ and the consequent VCN difference between two distinct AVH insulators $|\Delta C_K| = 1$. This is assumed in literatures for explaining the nontrivial interface modes in domain-wall systems [19-44]. The value of $\delta$ becomes notable with the increase of the gap size, as checked by the simulations for our concrete AVM systems [Fig. 1(d), bottom panel]. Physically, a sizeable $\delta$ makes the valley topology less well-defined, giving rise to a visible detachment of the interface modes from the bulk projections. More details can be seen in *Supplemental Materials* [60].

The subtle valley topology can be tracked from a 3D perspective of Weyl physics, if $\theta \in [-60°, 60°]$ is linearly mapped as an extra dimension of momentum, $k_z \in [-\pi, \pi]$. Given the linearly closed bandgap as $\theta \to \pm 30°$ (i.e., $\Delta_g \propto \Delta_\theta$, with $\Delta_\theta$ being an angular deviation from $\pm 30°$), the degenerate Dirac points (DPs) pinning at the inequivalent 2D BZ corners can be viewed as WPs in the synthetic 3D momentum space [61-64], around which the dispersions are linear in all directions [48-51]. Accordingly, the effective Hamiltonian in Eq. (1) turns out to be an anisotropic Weyl equation,
$$\delta H_w = v_D q_x \sigma_x + v_D q_y \sigma_y + v_z q_z \sigma_z, \quad (3)$$
where $v_z = \Delta_g/\Delta_\theta$ and $q_z = \Delta_\theta$. For clarity, in Fig. 1(b) we show the distribution of all WPs (color spheres) in the synthetic 3D BZ, where the adjacent WPs carry opposite charges $\pm 1$ and thus the total charge is zero, as required by the Nielsen-Ninomiya theorem. Physically, the WPs act as 3D monopoles of Berry curvature and enable many exotic properties such as gapless surface states along specific Weyl path and open Fermi arcs linking WPs [48-59]. Consequently, resorting to the strictly quantized WP charges, we can globally track the topological valley transport in individual AVMs with 3D Weyl topology. Below we start with the less explored single-crystal systems [8,9,42] and then generalize our study to the more popular domain-wall systems.

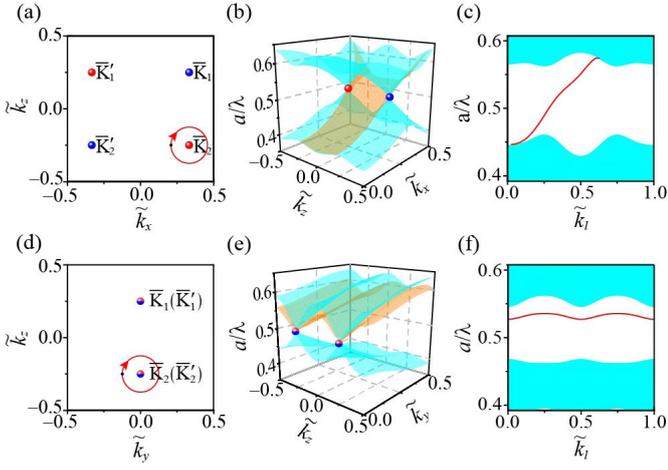

FIG. 2. Weyl path interpretation of the valley topology in rigid-boundary systems. (a) Synthetic SBZ projected along the $k_y$ direction, plotted with dimensionless momenta $\tilde{k}_x$ and $\tilde{k}_z$. The color spheres highlight the projections of synthetic WPs, and the red circle specifies a Weyl path encircling one projected WP. (b) Surface band (orange) plotted in the one half of SBZ. The blue surfaces sketch the boundaries of projected bulk bands. (c) Projected band structure (red line) along the clockwise Weyl path, which is characterized by the parameter $\tilde{k}_l = 0 \sim 1$ scaled with circumference. The blue-shaded areas are bulk projections. (d)-(f): Similar to (a)-(c), but for the projection along the $k_x$ direction. Note that the projected WPs at $K_1$ ($K_2$) and $K_1'$ ($K_2'$) are overlapped now. In contrast to (c), the surface states in (f) are no longer gapless since the Weyl path encircles a pair of oppositely-charged WPs.

*Gapless synthetic surface states.* Instead of directly characterizing WP charges, we identify the Weyl topology through detecting nontrivial surface states along specific Weyl paths [59]. Figure 2(a) sketches a synthetic surface BZ (SBZ) projected along the $k_y$ direction, in which the color spheres label the projected WPs. To unveil the global feature of the surface dispersion in the synthetic SBZ, we simulate the edge-projected spectra for a series of AVMs with different $\theta$ values, which are truncated in the $y$ direction with rigid boundaries. (As illustrated in Fig. 3(a) below, we consider the bottom edge only). For clarity, figure 2(b) shows the projected surface dispersion (orange surface) in one half of the SBZ; the data for the left one half can be inferred from time reversal. The nontrivial Weyl topology can be reflected in a Weyl path encircling one projected WP [Fig. 2(a)]. Geometrically, the closed loop can be viewed as the projection of a $k_y$-directed tube in the synthetic 3D BZ, through which a net Berry flux can be concluded from the nonzero WP charge inside the tube. Since the 2D subsystem defined on the tube carries a well-defined Chern number, topological surface states of specific chirality will emerge in the tube-projected loop according to the bulk-boundary correspondence [59]. In turn, we can identify the WP charge by inspecting the surface dispersion along any closed loop encircling the projected WP. As an example, figure 2(c) shows a synthetic surface spectrum extracted along a clockwise circular loop (of dimensionless radius 1/8) centered at $\bar{K}_2$. Clearly, it shows one gapless surface band (red line) connecting the upper and lower bulk bands. In particular, the surface band exhibits a positive slope, a faithful manifestation of the topological charge +1 for the WP locating at $K_2$. By contrast, we have also studied the surface dispersion projected along the $k_x$ direction [Figs. (2d)-(2f)], simulated for a series of AVMs with $y$-directed rigid boundaries. In this case, no gapless surface band can be observed for any closed loop, since each pair of oppositely-charged WPs are projected to the same point in SBZ. This is illustrated in Fig. 2(f) for a circular loop centered at $\bar{K}_2$ (overlapped with $\bar{K}_2'$). Note that the global connectivity of the surface band is irrelative to the selection of boundary details, in contrast to the remarkable boundary sensitivity of the edge states reflected in original 2D systems [60].

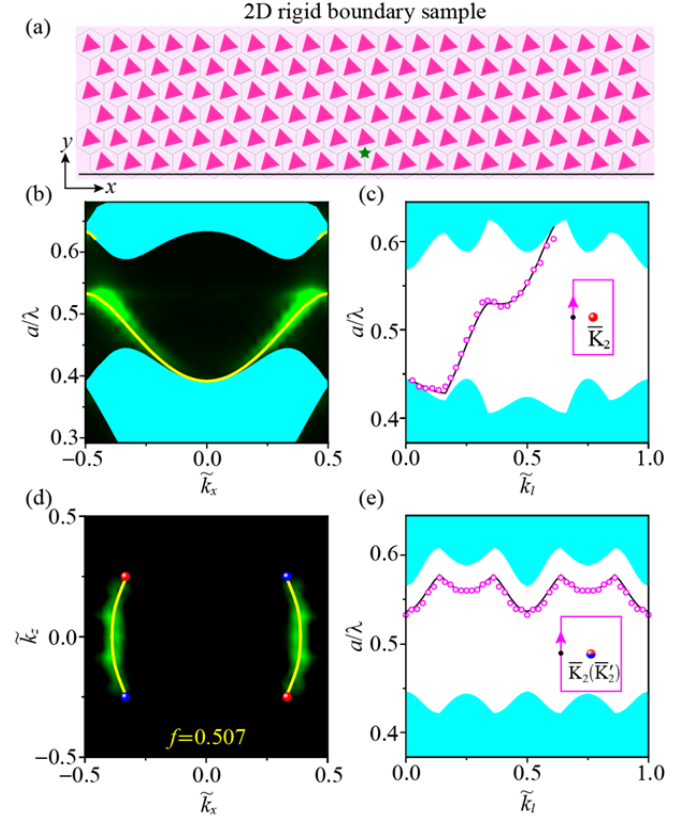

FIG. 3. Observations of the gapless chiral surface states and open surface arcs in synthetic momentum space. (a) Schematic of a rigid-boundary sample used for measuring 1D edge states along the $x$ direction. The green star highlights the sound source near the bottom edge (black line). (b) Measured (color scale) and simulated (yellow line) edge dispersions for the AVM with $\theta = 0°$. (c) Measured (open circles) and simulated (black line) surface bands along a rectangular loop centered at $\bar{K}_2$ in the $k_x$-$k_z$ SBZ (inset). Each circle corresponds to the peaked frequency for a given surface momentum. (d) Experimental evidence for the topological surface arcs at the synthetic WP frequency $f = 0.507$. (e) Similar to (c), but for a loop in the $k_y$-$k_z$ SBZ (inset).

*Acoustic experiments for rigid-boundary systems.* Acoustic experiments were performed to verify the synthetic Weyl topology. To map out the surface states in the synthetic SBZ, 1D edge states were detected for a series of 2D boundary-truncated samples at an angular step of 5°. Figure 3(a) sketches our experimental setup. To excite the edge states, a point-like sound source was located at the middle of the sample's bottom edge. We scanned the pressure distributions along the edge and attained the edge spectra by performing 1D Fourier transform (see more details in [60]). Figure 3(b) shows an example by $\theta = 0°$ (associated with a maximal band gap). As a direct manifestation of lacking bulk-boundary correspondence in individual sample, the measured edge band (bright color), which captures well the simulation result (yellow line), does not traverse the whole bulk gap. (It is consistent with the fact that no net Berry flux threads through a constant $k_z$ plane if considering from a viewpoint of the 3D Weyl crystal.) Repeating the measurements for the samples of different $\theta$, we obtain the global information of the surface states in the synthetic SBZ. Figure 3(c) shows the data (open circles) extracted along a rectangular loop (of dimensionless size $\frac{4}{15} \times \frac{1}{2}$) centered at $\overline{K}_2$. (Note that here the rectangular loop was used to reduce the number of 2D samples, since only one sample is required for each segment of horizontal path.) As expected, the surface dispersion features a single gapless chiral band with an overally positive slope. Without data presented here, a surface band of opposite chirality was identified for a loop encircling $\overline{K}_2'$ [60]. Another hallmark inherent in the Weyl physics is the presence of open surface arcs that link the oppositely-charged WPs. This was confirmed in Fig. 3(d), the synthetic surface dispersion extracted at the WP frequency, $f = 0.507$. Similar experiments were performed for the samples with $y$-directed edges. In contrast, no gapless surface band can be observed for any closed loop in the synthetic $k_y$-$k_z$ SBZ, because of the neutralized WPs with opposite charges. This is exemplified by a rectangular loop of size $\frac{2}{5} \times \frac{1}{2}$ in Fig. 3(e).

*Extensions to domain-wall systems.* The Weyl path interpretation of the valley physics can be extended to the widely explored domain-wall systems [19-44]. As sketched in the left panel of Fig. 4(a), the domain-wall system is formed by a pair of topologically distinct AVH insulators, whose scatterer orientations are related by the mirror parallel to the $x$ axis. In other words, the two AVMs can be described by coordinate systems of opposite chirality. Once introducing the additional $\theta$ degree of freedom, we obtain a 3D domain-wall system formed by two oppositely-charged Weyl crystals [Fig. 4(a), right panel]. Given the charge difference between them, in contrast to Fig. 2(c), each Weyl path enclosing $\overline{K}_2$ will host two gapless domain-wall states of the same chirality, according to the bulk-boundary correspondence. This is numerically exemplified in Fig. 4(b) by a circular path of dimensionless radius 1/8. Again, the conclusion does not rely on the structure details of the domain-wall, in contrast to the structure-sensitive interface dispersion exhibited in individual samples [60].

To confirm our theoretical prediction, we performed experiments like those of rigid-boundary systems, where the sound source was positioned in the middle of the domain-wall. Figure 4(c) exemplifies the measured 1D interface dispersion for a 2D domain-wall system formed with AVMs of $\theta_1 = 0°$ and $\theta_2 = 60°$. Comparing with the above single crystal case [Fig. 3(b)], the interface dispersion almost connects the upper and lower bulk projections, under the perfectly-matched interface configuration (which is widely used in exploring topological valley transport). The detachment of the interface modes from bulk will become notable for a system with a bigger bulk gap or a system with small Dirac velocity [60]. To experimentally identify the global connectivity of the interface states exhibited in synthetic Weyl topology, we repeated the interface measurements for a sequence of domain-wall samples and extracted the interface modes along a rectangular loop centered at $\overline{K}_2$. As expected, the data presented in Fig. 4(d) exhibits two gapless chiral interface bands. Furthermore, we extracted the interface states at the Weyl frequency. As shown in Fig. 4(e), the number of the surface arcs is doubled comparing with Fig. 3(d), as a consequence of the doubled Weyl charge difference in this system. All experimental data reproduce our simulations precisely.

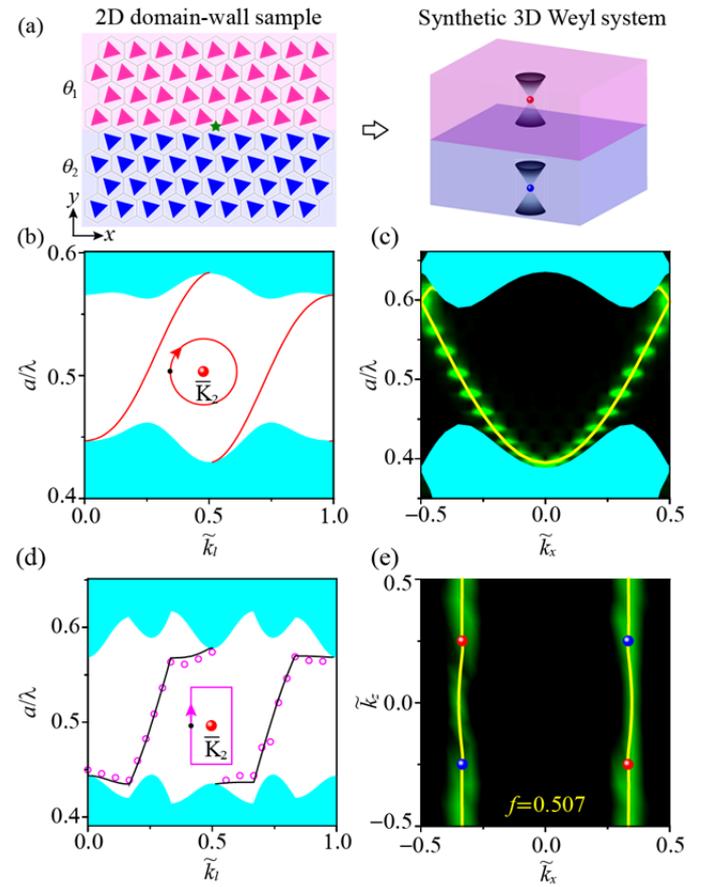

FIG. 4. Domain-wall systems. (a) Schematics of the 2D domain-wall system (left) and its synthetic extension to 3D (right). (b) Interface dispersion simulated along a circular Weyl path centered at $\overline{K}_2$ (inset). (c) Measured (color scale) and simulated (yellow line) interface dispersion for a domain-wall system formed by the AVMs with $\theta_1 = 0°$ and $\theta_2 = 60°$. (d) Measured (open circles) and simulated (black line) interface bands along a rectangular Weyl path centered at $\overline{K}_2$ (inset). (e) Measured and simulated isofrequency contour at the Weyl frequency.

*Conclusion.* We have proposed a 3D understanding for the 2D valley topology of metacrystals by constructing synthetic WPs. Both the rigid-boundary and domain-wall systems are considered in a unified framework. The theoretical prediction has been experimentally validated by measuring gapless edge/interface states along specific Weyl paths and open surface arcs linking oppositely-charged WPs. Our findings can be extended to the other classical (and even quantum) systems, if an appropriate parameter is selected to construct synthetic WPs. Novel applications would be enlightened by the deepened and global understanding to the subtle yet fundamental topological valley physics. Last but not the least, our study provides a concrete example that unveils intricate high-dimension band topology through easily controllable low-dimension physical systems [62-73]. In fact, similar methodology can also be employed to construct synthetic WPs with larger charges, e.g., by introducing a modulation degree of freedom to Kekulé lattices [74-78].

**Acknowledgements**
We thank M. Xiao, F. Zhang, and Z. Liu for fruitful discussions. This work is supported by the National Natural Science Foundation of China (Grant No. 11890701, 12004287, 12104346), the Young Top-Notch Talent for Ten Thousand Talent Program (2019-2022), the National Postdoctoral Program for Innovative Talents (Grant No. BX20200258), and the China Postdoctoral Science Foundation (Grant No. 2020M680107).